\documentstyle[prl,aps,twocolumn]{revtex}
\tighten

\begin{document}
\draft
\title{Photon-energy dissipation caused by an external electric circuit \\
in  ``virtual" photo-excitation processes}
\author{Akira Shimizu\cite{byline}}
\address{Institute of Physics, University of Tokyo, 3-8-1 Komaba, Tokyo 153,
Japan}
\author{Masamichi Yamanishi}
\address{Department of Physical Electronics, Hiroshima 
University, Higashi-hiroshima 724, Japan}
\date{\today}
\maketitle
\begin{abstract}
We consider generation of an electrical pulse by 
an optical pulse in the ``virtual excitation'' regime. 
The electronic system, which is  any electro-optic material including 
a quantum well structure biased by a dc electric field, 
is assumed to be coupled to an external circuit. 
It is found that the photon {\it frequency} is subject to an extra 
red shift in addition to the usual self-phase modulation, 
whereas the photon {\it number} is conserved. 
The Joule energy consumed in the external circuit 
is supplied only from the extra red shift. 
\end{abstract}

\narrowtext

Virtual excitation of electronic systems by optical fields 
has been attracting much attention recently \cite{Haug,Y,C,Y2,Yab}. 
Here, ``virtual'' means roughly that the photon energy is lower than 
the absorption edge by some detuning energy $\Delta$, 
so that photon absorption does not take place. 
More precisely, it means that 
the excitation 
takes place {\it adiabatically} so that no real transitions occur and 
the quantum-mechanical coherence is preserved.  
For a simple two-level system (with a long dephasing time), for example, 
the excitation will be ``virtual'' if 
\begin{equation}
(T_{\rm tr}/\hbar)^2 \gg (\mu {\cal E} / \Delta^2)^2, 
\label{adiabatic} \end{equation}
where $T_{\rm tr}$ is the transient time for which 
the envelope of the optical pulse changes appreciably, 
$\mu$ denotes the transition dipole moment of the two-level system, 
and $\cal E$ is the envelope of 
the electric-field amplitude of the optical pulse. 
The concept of the virtual excitation has been widely used, for example, to 
describe ultrafast nonlinear optical responses \cite{Haug,Y,C,Y2,Yab}. 
It was also shown that a `quantum non-demolition (QND)' measurement 
of the photon number $N$ ({\it i.e.}, measurement of $N$ 
 without changing its statistical distribution) 
is possible by the use of the virtual excitation of 
an electron interferometer \cite{as91}. 

On the other hand, many studies have recently been devoted to 
generation of an electrical pulse by excitation of a material by 
a short optical pulse \cite{Y,C,Y2,Yab,exp}. 
For the material, we can basically use 
any materials which possess finite electro-optic (EO) coefficient, 
$\chi^{(2)} \equiv \chi^{(2)}(0; -\omega, \omega)$. 
Of particular interest is 
 a quantum well structure (QWS) biased by a dc electric field \cite{Y,C}.
The dc field is applied to induce large $\chi^{(2)}$, which results in 
high efficiency for the generation of an electrical pulse.
In particular, it was suggested that 
the ultrafast response would be obtained by working in 
 the ``virtual excitation'' regime \cite{Y,C}. 
However, present understanding seems quite insufficient 
to investigate such a fancy combination of 
the idea of the electrical-pulse generation with the concept of 
the virtual excitation. 

In this paper, we raise and answer the following fundamental questions on  
the electrical-pulse generation by virtual photo-excitation: 
(i) What is the state of the optical pulse after it passes through the EO
material?
In particular, what is the photon energy and photon number?
(ii) What role is played by the external electric circuit in determining 
the photon state?
(iii) When the material is a biased QWS, 
what supplies the energy to the electrical pulse --- an 
external battery (which induce the dc bias field) or the optical field?
(iv) Is it possible to perform a QND measurement 
by monitoring the generated electrical pulse?

Let us start with a biased QWS.
We suppose that metallic contacts are deposited 
on both sides of a 
QWS sample 
in order to apply the static bias field $F_0$ ($ > 0$) by an external battery
$V_0$ (Fig.\ 1).  
The sample thus works as a capacitor, whose capacitance is 
$\epsilon L \equiv C_0$, where 
 $\epsilon$ denotes the 
linear dielectric constant at low frequencies. 
Exciton states of the QWS are strongly deformed by $F_0$, 
and each exciton acquires a large {\it static} dipole moment, $l$ \cite{Y,C}.

It is convenient to describe the exciton dynamics 
{\it in terms of such deformed states}. 
If, for simplicity, we look at the lowest-exciton state
only, 
an effective Hamiltonian in the optical field, 
${\cal E} \cos \omega t$, may then be written as  \cite{multilevel}
\begin{equation}
H= \varepsilon_x a^\dagger a
- \mu (a^\dagger + a) {\cal E} \cos \omega t
- l (F_P + F_1) a^\dagger a
\label{hamiltonian} \end{equation}
where $a^\dagger$ ($a$) and 
$\varepsilon_x$ denote the creation (annihilation) 
operator and the energy, respectively, of the deformed exciton state. 
When the detuning energy 
$\Delta \equiv \varepsilon_x - \hbar \omega$ satisfies Eq.\
(\ref{adiabatic}), 
the optical field excite the deformed excitons virtually, 
which induce the static electric field 
$F_P = - l \langle a^\dagger a \rangle / \epsilon_0$ \cite{units}, 
which is nonzero ($<0$) only in the well region \cite{Y,C}.
Since $l$ is large ($= 10^{1-2} \ e$\AA),  
$|F_P|$ becomes large, which results in 
a large EO coefficient $\chi^{(2)}$ \cite{Y,C,Y2,Yab}. 
On the other hand, 
to cancel out $F_P$, current $J$ is induced which alters the surface 
charge density of the metallic contacts 
from the equilibrium value $\sigma_0 \equiv \epsilon F_0$ 
into 
$\sigma = \sigma_0 + \sigma_1$, and 
 $\sigma_1$ generates the canceling field 
$F_1 = \sigma_1/\epsilon$. 
Therefore, the total dc field in the QWS is 
$F = F_0 + F_P \Theta(z) + F_1$, 
where $\Theta$ is a unit step function which is nonzero only in the well
region. 
The equation of motion of $\sigma_1$ may thus be 
\begin{equation} 
{d \sigma_1 \over d t} = - {\sigma_1 \over C_0 R} - \kappa {F_P \over R L}, 
\label{eqm}
\end{equation}
where $\kappa \equiv$ well thickness$/W$. 
The models of 
the previous work \cite{Y,C}, 
which assumed the absence of the external circuit, 
correspond to 
the limit of $R \rightarrow \infty$ of our model. 

It is seen that 
the optical field $\cal E$ interacts with the quantum-mechanical 
excitonic variables $a$, $a^\dagger$  
via the $\mu$ term in Eq.\ (\ref{hamiltonian}), 
and the excitonic variables interact with 
the classical surface charge $\sigma_1$ via the $F_P$ and $F_1$ 
terms in Eqs.\ (\ref{hamiltonian}) and (\ref{eqm}). 
The interesting point here is that 
{\it only} the motion of $\sigma_1$ suffers 
explicit dissipation; the $C_0 R$ term in Eq.\ (\ref{eqm}). 
We will show below that this dissipation eventually causes, 
through a chain of interactions, energy dissipation 
in the optical field.  

We first note that $F_0$ has not appeared explicitly in Eqs.
(\ref{hamiltonian}) 
and (\ref{eqm}): all effects of $F_0$ have been incorporated only 
in the deformed exciton state which defines $a$, $a^\dagger$, $\mu$, and $l$.
Consequently,  the external battery which produces $F_0$ 
supplies {\it no} net energy: the Joule energy $RJ^2$ must be supplied 
by something else --- the only possible supplier is the optical field.
The role of the battery (and $F_0$) is just to produce large $\chi^{(2)}$. 
The absence of energy supply from the battery 
 will be confirmed also in the following calculations. 

We are interested in the evolution of the optical field and energy flow. 
We here evaluate them to $O({\cal E}^2)$, because 
the analysis \cite{unp} which includes the third-order 
nonlinear effects shows that the second-order effects 
are essential \cite{third}. 
We also found \cite{unp} 
that concerning the quantities which we will discuss below 
the microscopic model of Eq.\ (\ref{hamiltonian}) gives the 
same results as a  
phenomenological model in which the excitonic (or electronic) system is 
phenomenologically treated as a transparent EO material. 
The only difference is that in the former model $\chi^{(2)}$ is 
obtained by solving Eq.\ (\ref{hamiltonian}), whereas in the 
latter $\chi^{(2)}$ is a given parameter.
The phenomenological model is therefore applicable to any 
transparent EO materials including the biased QWS. 
For this reason, we hereafter present our results in the language 
of the phenomenological  model: for example, $F_P$ is now 
\begin{equation}
F_P = - (\epsilon_0/\epsilon) \chi^{(2)} {\cal E}^2, 
\label{FP} \end{equation}
where $\chi^{(2)}$ is, as in the case of the biased QWS, 
the value of $\chi^{(2)}$ {\it in the presence of} $F_0$. 

To avoid inessential complexities, 
we assume that the light intensity is almost constant over the  
cross-section of the optical beam, and also that the cross-section  
agrees with that ($W \times W$) of the capacitor. 
In the propagating direction $x$, 
the optical pulse is assumed to have 
a portion (of length $c T$) of constant intensity
inbetween 
the initial and final transient portions of length $c T_{\rm tr}$. 
To focus on new phenomena only,  we assume that 
$T_{\rm tr} \ll C_0 R,\ T$; under this condition 
$\sigma$ does not change during the transients and thus 
what happens in the optical field of the transient portions is just 
the usual self-phase modulation, which is well-known and  
of no interest here. 
We therefore focus on the constant-intensity portion, and 
take $t=0$ as the time at which that portion begins to enter the capacitor. 
We further assume, for simplicity,  
that $L \ll c T/n$ ({\it i.e.}, ${\cal E} \approx$ constant in the capacitor), 
where $L$ is the length of the capacitor, 
$c$ the light velocity in vacuum,  and $n$ the refractive index. 

Under these conditions, Eq.\ (\ref{eqm}) can be easily solved to give
\begin{equation}
\sigma_1 = \left\{
\begin{array}{ll}
\kappa \epsilon |F_P| (1-e^{-t \over C_0 R}), 
& \quad (0 \leq t \leq T) \\
\kappa \epsilon |F_P| (1-e^{-T \over C_0 R}) e^{T-t \over C_0 R}, 
& \quad (T < t) \\
\end{array} \right.
\label{sigma}
\end{equation}
where $F_P$ is given by Eq.\ (\ref{FP}), 
and $\kappa$ is now 
$\kappa \equiv$ thickness of the EO material$/W$. 
Associated with the time-varying $\sigma_1$ is the 
current 
$J = {\partial \over \partial t} WL\sigma_1$, 
which generates the Joule heat in the resistance $R$ \cite{R}; 
\begin{eqnarray}
U_R & = & \int_{-\infty}^\infty RJ^2 dt \nonumber \\
& = & (\kappa WL \epsilon_0 \chi^{(2)} {\cal E}^2)^2
      (1-e^{-T/C_0 R})/C_0.   
\label{Joule} \end{eqnarray}
Let us find out the supplier of this energy --- the battery or 
the optical field?
The work done by the battery is 
\begin{equation}
U_{V_0} = \int_{-\infty}^\infty V_0 J dt 
= V_0 WL[ \sigma_1(\infty) - \sigma_1(0)], 
\end{equation}
which is zero because, as seen from Eq.\ (\ref{sigma}), 
 $\sigma_1(\infty) = \sigma_1(0)=0$. 
That is, the battery does not supply net 
energy at all, in agreement with the observation we have drawn 
above from the microscopic model. 
Therefore, the only possible supplier 
of the Joule energy is the optical field, 
the evolution of which we will investigate now. 

We note that the dc field $F$ in the EO material varies from
$F_0$ (for $t < -T_{\rm tr}$) to $F_0+F_P+F_1$ (for $0 \leq t \leq T$) 
and then to $F_0+F_1$ (for $T+T_{\rm tr} < t$), where 
$F_1=\sigma_1/\epsilon$ also varies according to 
Eq.\ (\ref{sigma}). 
The  time-dependent $F$ 
produces the time-dependent change of 
the refractive index $n$:
\begin{equation}
\delta n = \chi^{(2)}(\omega; \omega, 0) (F-F_0) / 2 n, 
\label{dn1} \end{equation} 
where $n$ and $\chi^{(2)}(\omega; \omega, 0)$ denote their values 
{\it in the presence of} $F_0$. 
With the help of the symmetric relation, 
$
\chi^{(2)}(\omega; \omega, 0)= 4 \chi^{(2)}(0; -\omega, \omega),
$
Eqs.\ (\ref{sigma}) and (\ref{dn1}) yield, for $0 \leq t \leq T$,  
\begin{equation}
\delta n = -(2 \epsilon_0/n \epsilon) |\chi^{(2)}|^2 {\cal E}^2 
[1-\kappa(1-e^{-t/C_0 R})].
\label{dn2} \end{equation} 
By this time-dependent $\delta n$, the optical field undergoes 
a frequency shift (chirping) \cite{unp}. 
When $L \ll C_0 R c/n$, 
the shift is simply given by
\begin{eqnarray}
\delta \omega 
& = & - {\partial \over \partial t} {\omega  L \kappa \delta n \over c}
\nonumber \\
& = & - {2 \epsilon_0 \omega \kappa^2 |\chi^{(2)}|^2 {\cal E}^2 L
\over 
n \epsilon c C_0 R}
e^{-t/C_0R}. 
\label{do} 
\end{eqnarray}
Here $\kappa$ in the first line has appeared because 
$\delta n$ is large only in the EO material. 

We find that 
(i) the optical field undergoes a red shift, 
(ii) the shift approaches zero in both limits of
$R \rightarrow \infty$ and $R \rightarrow 0$, 
and 
(iii) the shift becomes maximum at the beginning ($0 \leq t \ll C_0 R$) 
of the constant-intensity portion of the optical pulse, 
and decays exponentially after that. 

This shift is a kind of a self phase modulation (SPM) 
process in the sense that 
the shift is driven by the optical field {\it itself}. 
However, it is totally different from the {\it usual}  
SPM, which generally occurs when 
an optical pulse passes through a nonlinear medium. 
To distinguish between the two, we 
hereafter call the shift of Eq.\ (\ref{do}) 
the `extra shift' or `extra red shift.' 
Major differences are; 
(a) In contrast to Eq.\ (\ref{do}), the usual SPM is basically 
independent of the external circuit --- it occurs, for example, 
even when $R \rightarrow \infty$. 
(b) The {\it total} energy of the optical pulse is conserved in the usual 
SPM process (because the frequency shifts occur in the opposite 
directions at the initial and final transients), whereas 
the extra shift results in loss of the total energy (Eq.\ (\ref{ERS}) below).
(c) The usual SPM is approximately instantaneous 
(delay $\approx$ response time of nonlinear processes), 
whereas the extra shift 
occurs with a considerable delay ($\sim C_0 R$)
--- the shift takes place during $0 \lesssim t \lesssim C_0 R$ 
in order to compensate for $F_P$ which is established at $t=0$.  
These differences arise because the extra shift is a property of the 
{\it coupled} system of a nonlinear EO material and an external electric 
circuit, whereas the usual SPM is a property of a nonlinear material only. 

In terms of the microscopic model, 
the physics of the extra shift may be understood as follows:
Photons (virtually) excite excitons of energy 
$\varepsilon_x - l (F_P + F_1)$, 
and the excited excitons will emit photons subsequently. 
Here, the energy of the excitons is decreasing as $t$ goes by because 
$F_1 = \sigma_1/\epsilon$ is increasing according to Eq.\ (\ref{sigma}). 
As a result, the emitted photons have lower energies than 
the (virtually) absorbed photons. Hence the red shift, and 
its magnitude decays exponentially 
with the same decay constant as that of $\sigma_1$. 
 
The magnitude of the extra shift, Eq.\ (\ref{do}), depends on many 
material and structural parameters. 
For example, 
for a 100-\AA well/ 100-\AA barrier 
multiple quantum well structure for which $\kappa \approx 1/2$, 
the red shift is estimated to be of the order of $10^2 L$ MHz when 
$I \sim 10^2$ MW/cm$^2$, $F_0 \sim10^2$ kV/cm, 
$T \sim C_0 R \sim 1$ ps, 
$\Delta =$ 10 meV, and $L$ here is measured in $\mu$m, 

Our next task is to find out the energy flow.  
To do this, we for the moment assume that the photon number is 
conserved (this assumption will be justified soon). 
In this case, the loss of the light intensity
 $I= \epsilon_0 c n {\cal E}^2 /2$ is given by  
$\delta I = I \delta \omega / \omega$. 
Therefore, the loss of the total photon energy due to the extra 
red shift is
\begin{eqnarray}
U_{ERS} & = & \int_0^T \left| W^2 \delta I \right| dt \nonumber \\
& = & W^2 I L \kappa [\delta n(0)-\delta n(T)]/c.
\label{ERS}
\end{eqnarray}
Inserting Eq.\ (\ref{dn2}) and $C_0 = \epsilon L$, 
and comparing with Eq.\ (\ref{Joule}), 
we find that $U_{ERS} = U_R$. 
Therefore, {\it all the Joule energy is supplied by 
the extra red shift of the optical field. } 
It also shows that {\it the photon number $N$ is 
conserved}, because if it were lost then $U_{ERS} < U_R$. 
This indicates that the present photon-energy dissipation 
{\it cannot be described as a simple dephasing process}, 
which is accompanied by loss of $N$.  

We have thus found that when an optical pulse excites 
the electronic system of Fig.\ 1 `virtually' 
(in the sense that Eq.\ (\ref{adiabatic}) 
is satisfied) then the final state of 
the optical pulse 
has the {\it same} number of photons as the initial state. 
However, the frequencies 
of photons are lowered, 
which are consumed to generate the Joule heat in the external circuit.   
This is in a marked contrast to the 
virtual photo-excitation of an electron interferometer 
which was discussed 
in \cite{as91}, where it was shown that 
{\it both} the number and frequencies of photons are conserved. 
Since we can estimate the photon number $N$ by  measuring 
the interference currents, 
the electron interferometer works as a QND photo-detector \cite{as91}. 
We can estimate $n$ also in the present case by, say,  
monitoring the voltage drop across the resistance. 
Can we call it a QND measurement? The answer clearly 
depends on the definition of the QND measurement. 
Kitagawa \cite{kita} proposed to accept it as 
a QND measurement in a broad sense. 
To perform a QND measurement in the narrow sense 
({\it i.e.}, both $N$ and frequencies are conserved), 
one may use the scheme of Ref.\ \cite{Y2}, in which
the generated voltage modulates 
an electron interference current in  
an electro-static Aharonov-Bohm interferometer.  

We have thus found that when you try to get information on photons 
through virtual photo-excitation of an electronic system, 
the photon energy will or will not be conserved depending on 
the detailed structures of the electronic system {\it and} the external 
circuit (although the circuit is {\it not} directly 
connected to the optical field). 
These findings shed light on the theory of measurement 
of photons using electronic systems. 

Finally, let us comment on the case 
in which 
the external circuit of Fig.\ 1 is a transmission line or something like
that, 
which has a complex impedance $Z$ rather than the pure resistance $R$. 
The extension of the present theory to such a general case is straightforward
---
all we have to do is to modify the last term of Eq.\ (\ref{eqm}). 
We would then observe,  
for example, that the extra shift would exhibit an oscillatory behavior 
which is superposed on the exponential decay. 
However, the main conclusions  
do not change because, for example,  the irrelevance of the battery 
in the energy consumption processes relies on the fact that 
$F_0$ does not appear explicitly 
in Eqs.\ (\ref{hamiltonian}) and (\ref{eqm}) --- this fact 
remains true when we modify the last term of Eq.\ (\ref{eqm}). 
For this reason, we believe that the present paper 
has revealed bare essentials of 
the electrical-pulse generation by the virtual photo-excitation. 

To summarize, 
we have considered electrical-pulse generation 
in the ``virtual excitation'' regime. 
The electronic system 
is any electro-optic material including 
a quantum well structure (QWS) biased by a dc electric field, 
which is applied to induce large $\chi^{(2)}$. 
The energy transfer is analyzed when the electronic system is 
coupled to an external circuit. 
It is found that 
the photon {\it frequency} is subject to an extra red shift in addition to
the usual 
self-phase modulation, 
whereas the photon {\it number} is conserved. 
The extra red shift approaches zero in both limits of zero and  
infinite impedance of the circuit. 
It is also shown that 
an external battery,  which produces the dc bias field in the QWS, 
supplies {\it no} net energy, and 
the Joule energy consumed in the external circuit 
is supplied only from the extra red shift 
of the optical field.

\begin{figure}
\caption{
A schematic diagram of the system under consideration. 
An optical pulse goes into a capacitor, the center region of 
which is made of a biased QWS 
or another EO material. 
An electrical pulse is generated in the center region, and 
the current $J$ flows in the external circuit. 
We find that the optical pulse, after it passes through the capacitor, 
is subject to an extra red shift, in addition to the usual self phase
modulation 
which occurs in the initial and final transients. 
}
\end{figure}


\begin{references}
\bibitem{units} 
We are working in the SI units, and $\epsilon_0$ 
denotes the vacuum dielectric constant.
\bibitem{third}
The third-order effects, for example, modify the decay constant $C_0R$ 
in Eqs.\ (\ref{sigma})-(\ref{ERS}). 
This, however, does not alter the main conclusions such as 
$U_R = U_{ERS}$ and $U_{V_0} = 0$.
\bibitem{R}
We include in $R$ the internal resistance of the external battery.
\end{references}
\end{document}